\pdfoutput=1
\documentclass{emulateapj}
\usepackage{natbib}
\newcommand{\Msun}{$M_{\odot}$}
\newcommand{\Lsun}{$L_{\odot}$}

\slugcomment{Accepted by ApJ}

\shorttitle{A large mass of cold dust in the ejecta of Supernova 1987A}
\shortauthors{Matsuura et al.}

\begin{document}


\title{A stubbornly large mass of cold dust  in  the ejecta of Supernova 1987A}


\author{
M. Matsuura\altaffilmark{1}, 
E. Dwek\altaffilmark{2}, 
M. J. Barlow\altaffilmark{1}, 
B. Babler\altaffilmark{3}, 
M. Baes\altaffilmark{4}, 
M. Meixner\altaffilmark{5},
Jos{\'e} Cernicharo\altaffilmark{6}, 
Geoff C. Clayton\altaffilmark{7}, 
L. Dunne\altaffilmark{8, 9}, 
C. Fransson\altaffilmark{10}, 
Jacopo Fritz\altaffilmark{4},
Walter Gear\altaffilmark{11}, 
H. L. Gomez\altaffilmark{11}, 
M.A.T. Groenewegen\altaffilmark{12}, 
R. Indebetouw\altaffilmark{13,14},
R.J. Ivison\altaffilmark{14,15}, 
A. Jerkstrand\altaffilmark{16},
V. Lebouteiller\altaffilmark{17},
T. L. Lim\altaffilmark{18}, 
P. Lundqvist\altaffilmark{10}, 
C.P. Pearson \altaffilmark{18},
J Roman-Duval\altaffilmark{5}, 
P. Royer\altaffilmark{19}, 
Lister Staveley-Smith\altaffilmark{20,21}, 
B.M. Swinyard\altaffilmark{1,18}
P.A.M. van Hoof\altaffilmark{12}, 
J.Th. van Loon\altaffilmark{22},
Joris Verstappen\altaffilmark{4 , 23}, 
Roger Wesson\altaffilmark{24}, 
Giovanna Zanardo\altaffilmark{20}, 
Joris A.D.L. Blommaert\altaffilmark{19,  25}, 
Leen Decin\altaffilmark{19}, 
W.T. Reach\altaffilmark{26}, 
George Sonneborn\altaffilmark{2}, 
Griet C. Van de Steene\altaffilmark{12}, 
Jeremy A. Yates\altaffilmark{1}}
\altaffiltext{*}{{\em Herschel} is an ESA space observatory with science instruments provided by European-led Principal Investigator consortia and with important participation from NASA.
PACS has been developed by a consortium of institutes led by MPE (Germany) and including UVIE (Austria); KU Leuven, CSL, IMEC (Belgium); CEA, LAM (France); MPIA (Germany); INAF-IFSI/OAA/OAP/OAT, LENS, SISSA (Italy); IAC (Spain). This development has been supported by the funding agencies BMVIT (Austria), ESA-PRODEX (Belgium), CEA/CNES (France), DLR (Germany), ASI/INAF (Italy), and CICYT/MCYT (Spain).
SPIRE has been developed by a consortium of institutes led by Cardiff University (UK) and including Univ. Lethbridge (Canada); NAOC (China); CEA, LAM (France); IFSI, Univ. Padua (Italy); IAC (Spain); Stockholm Observatory (Sweden); Imperial College London, RAL, UCL-MSSL, UKATC, Univ. Sussex (UK); and Caltech, JPL, NHSC, Univ. Colorado (USA). This development has been supported by national funding agencies: CSA (Canada); NAOC (China); CEA, CNES, CNRS (France); ASI (Italy); MCINN (Spain); SNSB (Sweden); STFC and UKSA (UK); and NASA (USA).
}
\altaffiltext{1}{Department of Physics and Astronomy, University College London, Gower Street, London WC1E 6BT, UK}
%\email{(e-mail: m.m.) mikako@star.ucl.ac.uk}
\altaffiltext{2}{Observational Cosmology Laboratory Code 665, NASA Goddard Space Flight Center, Greenbelt, MD 20771, USA}
\altaffiltext{3}{Department of Astronomy, 475 North Charter St., University of Wisconsin, Madison, WI 53706, USA}
\altaffiltext{4}{Sterrenkundig Observatorium, Universiteit Gent, Krijgslaan 281 S9, B-9000 Gent, Belgium}
\altaffiltext{5}{Space Telescope Science Institute, 3700 San Martin Drive, Baltimore, MD 21218, USA}
\altaffiltext{6}{Departamento de Astrof\'{i}sica, Centro de Astrobiolog\'{i}a, CSIC-INTA, Ctra. de Torrej\'{o}n a Ajalvir km 4, E-28850 Madrid, Spain}
\altaffiltext{7}{Department of Physics \& Astronomy, Louisiana State University, Baton Rouge, LA 70803, USA}
\altaffiltext{8}{Department of Physics and Astronomy, University of Canterbury, Private Bag 4800, Christchurch 8140, New Zealand}
\altaffiltext{9}{SUPA, Institute for Astronomy, University of Edinburgh, Blackford Hill, Edinburgh EH9 3HJ, UK}
\altaffiltext{10}{The Oskar Klein Centre, Department of Astronomy, Stockholm University, Albanova, SE-10691 Stockholm, Sweden}
\altaffiltext{11}{School of Physics and Astronomy, Cardiff University, Cardiff CF24 3AA, UK}
\altaffiltext{12}{Koninklijke Sterrenwacht van Belgi\"{e}, Ringlaan 3, 1180, Brussel, Belgium}
\altaffiltext{13}{Department of Astronomy, University of Virginia, PO Box 400325, Charlottesville, VA 22904, USA}
\altaffiltext{14}{National Radio Astronomy Observatory, 520 Edgemont Rd, Charlottesville, VA 22903, USA}
\altaffiltext{15}{European Southern Observatory, Karl-Schwarzschild-Strasse 2, D-85748 Garching, Germany}
\altaffiltext{16}{Astrophysics Research Centre, School of Mathematics and Physics, Queen's University Belfast, Belfast BT7 1NN, UK}
\altaffiltext{17}{AIM, CEA/Saclay, L'Orme des Merisiers, 91191, Gif-sur-Yvette, France}
\altaffiltext{18}{RAL Space, Rutherford Appleton Laboratory, Chilton, Didcot, Oxfordshire, OX11 0QX, UK}
\altaffiltext{19}{Instituut voor Sterrenkunde, Katholieke Universiteit Leuven, Celestijnenlaan 200D, 3001 Leuven, Belgium}
\altaffiltext{20}{International Centre for Radio Astronomy Research (ICRAR), M468, The University of Western Australia, Crawley, WA 6009, Australia}
\altaffiltext{21}{Australian Research Council, Centre of Excellence for All-sky Astrophysics (CAASTRO)}
\altaffiltext{22}{Lennard Jones Laboratories, Keele University, ST5 5BG, UK}
\altaffiltext{23}{Kapteyn Astronomical Institute, P.O. Box 800, 9700 AV Groningen, the Netherlands}
\altaffiltext{24}{European Southern Observatory, Alonso de C\'{o}rdova 3107, 19001 Casilla, Santiago, Chile}
\altaffiltext{25}{Astronomy and Astrophysics Research Group, Department of Physics and Astrophysics, Vrije Universiteit Brussel, Pleinlaan 2, 1050 Brussels, Belgium}
\altaffiltext{26}{Stratospheric Observatory for Infrared Astronomy, Universities Space Research Association, MS 232-12, NASA/Ames Research Center, Moffett Field, CA 94035, United States}


\begin{abstract}

We present new {\em Herschel} photometric and spectroscopic observations 
of Supernova 1987A, carried out in 2012. Our dedicated photometric 
measurements provide new 70\,$\mu$m data and improved imaging quality at 
100 and 160\,$\mu$m compared to previous observations in 2010. Our {\em 
Herschel} spectra show only weak CO line emission, and provide an upper 
limit for the 63\,$\mu$m [O~{\sc i}] line flux, eliminating the 
possibility that line contaminations distort the previously estimated dust 
mass. The far-infrared spectral energy distribution (SED) is well fitted 
by thermal emission from cold dust. The newly measured 70\,$\mu$m flux 
constrains the dust temperature, limiting it to nearly a single 
temperature. The far-infrared emission can be fitted by 
0.5$\pm$0.1\,\Msun\ of amorphous carbon, about a factor of two larger than 
the current nucleosynthetic mass prediction for carbon. The observation of 
SiO molecules at early and late phases suggests that silicates may 
also have formed and we could fit the SED with a 
combination of 0.3\,\Msun\, of amorphous carbon and 0.5\,\Msun\, of 
silicates, totalling 0.8\,\Msun\, of dust. Our analysis thus supports the 
presence of a large dust reservoir in the ejecta of SN~1987A. The 
inferred dust mass suggests that supernovae can be an important source of 
dust in the interstellar medium, from local to high-redshift galaxies.
 
\end{abstract}


\section{Introduction}

The explosion of Supernova (SN) 1987A 
in the Large Magellanic Cloud (LMC)
was detected on 23 February 1987 \citep{1987IAUC.4316....1K}.
SN 1987A has since provided a unique opportunity to study the evolution of SN ejecta and SN remnants.

One of the early discoveries was the detection of thermal emission from 
dust, believed to have formed in the ejecta. The emission, appearing at 
mid-infrared wavelengths, probably began at about day 260, increasing in
flux to day 1316
 \citep{Wooden:1993p29432, Bouchet:1991bt}.
The emission was attributed to $\sim 10^{-4}$\,\Msun\, of dust.
About 23 years later, the {\em Herschel} Space Observatory detected 
thermal dust emission at 100--350\,$\mu$m \citep{Matsuura:2011ij}, with 
ALMA resolved images confirming that the far-infrared emitting dust was 
located in the ejecta \citep{Indebetouw:2013vv}. The {\em Herschel}-based 
dust mass was three orders of magnitude larger (0.4--0.7\,\Msun) than 
previously reported. This surprisingly large dust mass triggered debates 
about the nature of the far-infrared emission, not only because it was far 
larger than the measurements made at much earlier epochs, but also because 
it was much larger than typical dust masses that had been deduced from 
{\em Spitzer} mid-infrared observations of other core-collapse SNe during 
their first three years after outburst, $10^{-6}$--$10^{-4}$\,\Msun\, 
\citep[e.g.][]{Gall:2011hr}, although {\em Herschel} 
far-infrared observations of two historical SN remnants, Cassiopeia A and 
the Crab Nebula \citep[e.g.][]{Barlow:2010p29287, Gomez:2012fm}
found significantly larger dust masses ($\geq$0.1~M$_\odot$). Possible 
ways to reduce the dust mass derived for SN~1987A have been proposed, 
including line contamination in the {\em Herschel} filter band-passes. 
Because the initial detection was made from fast-scan observations 
(leading to spatial under-sampling for PACS) during the {\em Herschel} 
HERITAGE survey of the LMC \citep{Meixner:2013kr}, and because possible 
line contamination needed to be evaluated, we obtained {\em Herschel} 
dedicated PACS and SPIRE observations of SN 1987A in 2012 using both their 
photometric and spectroscopic modes, which we report here.

\section{Observations and data reduction}

\subsection{PACS and SPIRE imaging}\label{photometry}

The {\it Herschel Space Observatory} \citep{Pilbratt:2010p29312} detected 
SN\,1987A at far-infrared and submillimeter wavelengths in 2010 
\citep{Matsuura:2011ij}, as part of the HERschel Inventory of the Agents 
of Galaxy Evolution \citep[HERITAGE; ][]{Meixner:2013kr}. The survey used 
five filter bands from 100--500\,$\mu$m and SN\,1987A was detected in the 
four bands from 100--350\,$\mu$m. {\em Herschel} scanned SN 1987A on 30 
April and 5 August 2010 (days 8467 and 8564 after the explosion). We adopt 
the SPIRE 250 and 350\,$\mu$m fluxes from the HERITAGE point source 
catalogue \citep{Meixner:2013kr}. The HERITAGE PACS images were affected 
by residual striping due to $1/f$ noise \citep{Meixner:2013kr} and 
dedicated stripe removal procedures were adopted for the image 
reconstruction for SN\,1987A.

Dedicated {\em Herschel} photometric observations were carried out in 
2012, acquired as part of a guaranteed time observing programme 
(GT2$_-$mbaes$_-$3). The PACS \citep{Poglitsch:2010bm} images (OBSID 
1342237428, 1342237429, 1342237430 and 1342237431) were acquired on 2012 
January 13th (UT), corresponding to day 9090 after the explosion. The 
large scan map mode was used. The observing sequence was composed of two 
observations, one to obtain 100 and 160\,$\mu$m cross-scan images with 
2$\times$1295-secs duration, and the other to obtain 70 and 160\,$\mu$m 
cross-scan images with 2$\times$2245-secs duration. The FWHMs of the point 
spread functions (PSFs) were 5.46$\times$5.76, 6.69$\times$6.89 and 
10.65$\times$12.13\,arcsec$^{2}$ for the PACS\,70, PACS\,100 and PACS\,160 
bands, respectively (PACS observer's 
manual\footnote{http://herschel.esac.esa.int/Docs/PACS/html/pacs$_{-}$om.html}). 
The absolute flux calibration uncertainties of the PACS photometer are 
estimated to be 3\,\% for the 70 and 100\,$\mu$m bands, and 5\,\% for the 
160\,$\mu$m band.

The SPIRE \citep{Griffin:2010hz} images of SN\,1987A (OBSID 1342239283) 
were obtained using the large scan map mode on 2012 February 14th (UT), 
corresponding to an epoch of 9122 days. With an integration time of 
2553\,sec, we simultaneously obtained 10\,arcmin$\times$10\,arcmin images 
at 250, 350, and 500\,$\mu$m. The FWHMs of the beams were 18.2, 24.9 and 
36.3\,arcsec at 250, 350 and 500\,$\mu$m, respectively 
\citep{Griffin:2013kq}. %and the final images were re-sampled to have 5, 8 
and 11\,arcsec pixel scales at 250, 350 and 500\,$\mu$m, respectively. 
The absolute flux calibration errors were estimated to be 5\,\% 
\citep{Bendo:uv}. The colour correction factors are less than 1\,\%, so we 
ignore them.

\begin{figure*}
\centering
%\rotatebox{90}{ 
\begin{minipage} {13cm} 
\resizebox{\hsize}{!}{\includegraphics{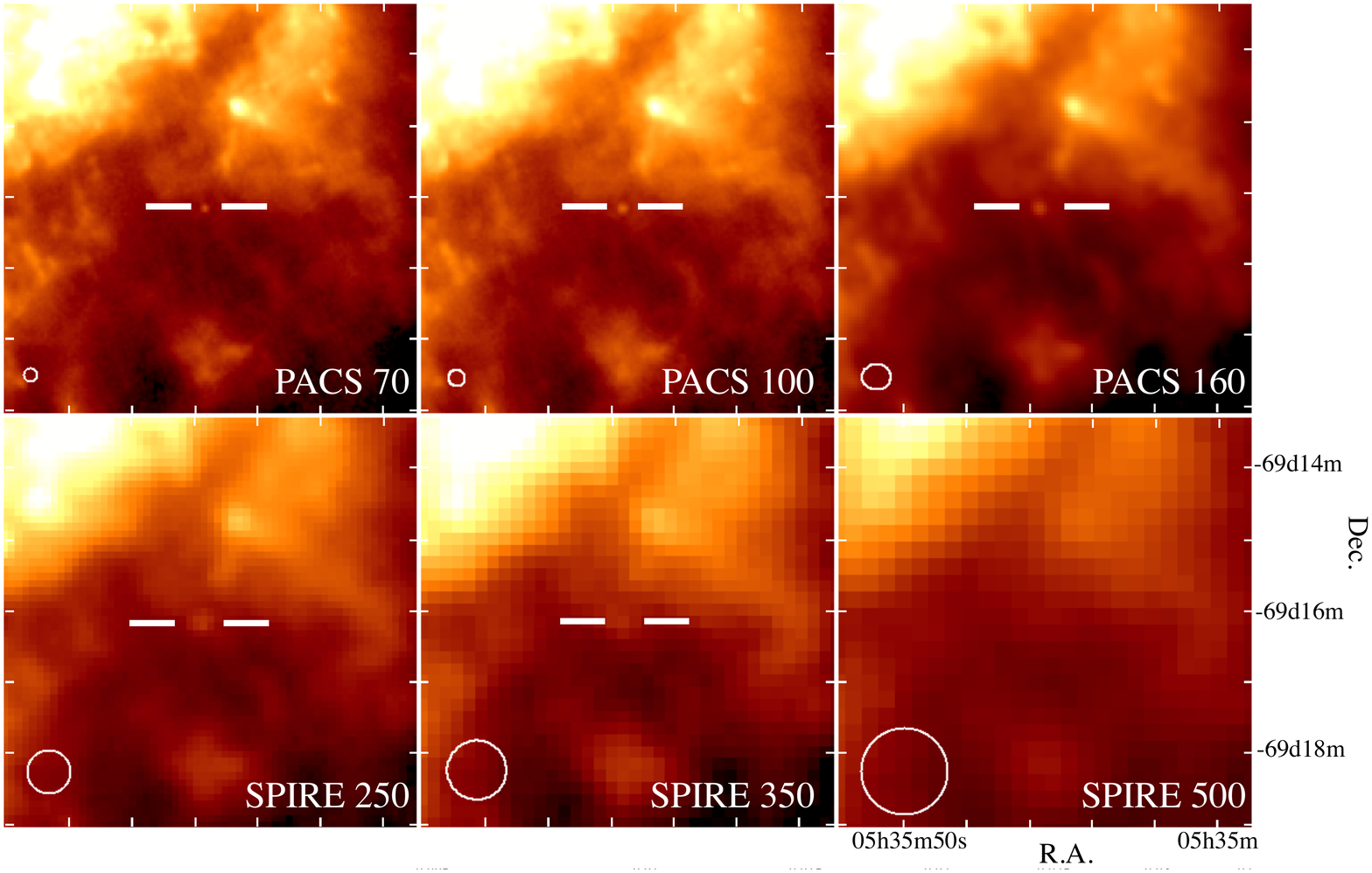}}
\end{minipage}
%}
%\resizebox{\hsize}{!}{\includegraphics{ds9.ps}}
\caption{The PACS and SPIRE images of SN\,1987A and its surroundings
 from the dedicated  observations taken in 2012.
The supernova is found as a point source. The white circles show the size of the PSFs. 
\label{fig-image}}
\end{figure*}

Fig.\,\ref{fig-image} shows the PACS and SPIRE images of SN 1987A and its 
surroundings from the {\em Herschel} 2012 observations. SN\,1987A was 
detected as a point source from 70--350\,$\mu$m, whereas it was not 
clearly detected in the SPIRE 500 image, because it is diluted by LMC 
interstellar medium (ISM) emission in the large 500\,$\mu$m beam.

The {\sc idl} PSF-fitting code, {\em starfinder} 
\citep{Diolaiti:2000p29450} was used to obtain point source photometric 
measurements; details are given by \citet{Meixner:2013kr}. The 
uncertainties in the fluxes include the error maps from the pipeline, the 
uncertainties in the absolute flux calibration, the fluctuations in the 
sky level, as well as {\em starfinder}'s fitting uncertainties.

Table\,\ref{table-flux} lists the measured {\it Herschel} fluxes, which 
are consistent between 2010 and 2012 within the uncertainties. The pointed 
observations have reduced uncertainties for the PACS 100 and 160\,$\mu$m 
fluxes, because the optimised imaging for a point source provides higher 
sampling rates by a factor of three, improving the overall image quality 
in the PACS bands. The SPIRE 250 and 350\,$\mu$m fluxes are consistent 
between the HERITAGE and the pointed observations within the 1 $\sigma$ 
uncertainties, as the uncertainties are dominated by uncertainties of the 
sky estimates.

\begin{figure*}  %figure 2
\centering
\resizebox{0.8\hsize}{!}{\includegraphics{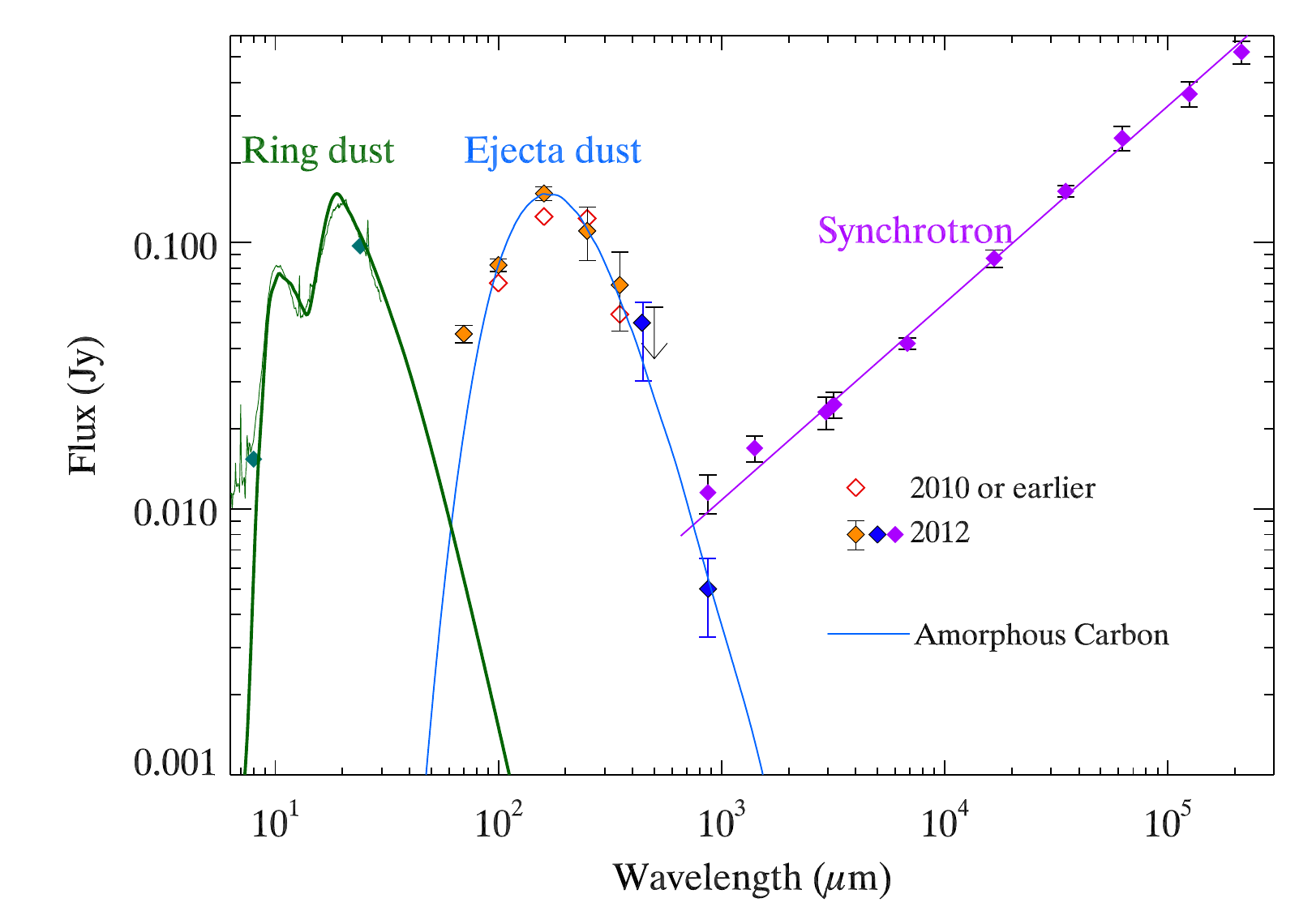}}
\caption{The 2010 and 2012 SED of SN1987A, showing two distinct components of
thermal dust emission: warm dust from the equatorial ring and cold dust 
from the ejecta. Additionally, synchrotron radiation from the ring is 
detected at millimeter wavelengths and longer, showing a power law 
frequency dependence \citep[][Zanardo et al. in 
preparation]{StaveleySmith:2014dw, Zanardo:2014}. Two dust models are 
plotted - the parameters for the cold ejecta dust can be found in 
Table\,\ref{table-dust} (model b), while the silicate fit to the ring dust 
is plotted as a thick green line (see text).
\label{fig-sed}
}
\end{figure*}

Figure\,\ref{fig-sed} presents the spectral energy distribution (SED) of 
SN\,1987A from infrared to millimetre wavelengths. In addition to our {\em 
Herschel} measurements, we assembled additional flux measurements from the 
literature. The day~9090 mid-infrared photometric points were extrapolated 
from measurements made on day 7983, adopting the empirical power-law fit 
of \citet{Dwek:2010kv} to the flux increase with time. The estimated day 
9090 fluxes are 15.4$\pm$0.4\,mJy and 97$\pm$5\,mJy at 8.0 and 24\,$\mu$m, 
respectively. The mid-infrared fluxes have been shown to be due to thermal 
emission from silicate dust grains located in the equatorial ring 
\citep{Bouchet:2006p2168}. We fitted these fluxes with silicate dust 
emission \citep{Weingartner:2001p3411}; its day 9090 parameters are a dust 
temperature of 187\,K and a dust mass of $1.6\times10^{-5}$\,\Msun. The 
{\it Spitzer} IRS spectrum at day 7965 \citep{Dwek:2010kv}, which was 
estimated to correspond to a dust temperature of T$\sim$175\,K, was scaled 
by $B_{\nu}(T=187~K)/B_{\nu}(T=175~K)$, where $B_{\nu}$ is the Planck 
function, and is plotted as the thin green line in Figure\,\ref{fig-sed}.
The millimetre-wavelength 
fluxes are fitted by a synchrotron radiation spectrum; these fluxes 
increase in time and the fluxes were scaled to day~9090 from recent ATCA 
observations \citep[][Zanardo et al. in preparation]{Zanardo:2014}, or 
scaled to the Herschel epoch via exponential fitting of the flux densities 
from day $\sim$8000 \citep[][Zanardo et al. in 
preparation]{Zanardo:2010p29272, StaveleySmith:2014dw}.

%====================
%====================

%_________________________________________________________________
\begin{table}
% \centering
  \caption{   The measured far-infrared and submillimeter flux densities of SN 1987A \label{table-flux}}
\begin{center}
 \begin{tabular}{l r@{$\pm$}l r@{$\pm$}l  l}
\hline
Band name  & \multicolumn{4}{c}{Flux (mJy)}    & Reference \\ 
 &  \multicolumn{2}{c}{HERITAGE} &  \multicolumn{2}{c}{Follow-up}  \\ 
  & \multicolumn{2}{c}{2010} & \multicolumn{2}{c}{2012} & \\
 \hline
PACS   70\,$\mu$m  & \multicolumn{2}{c}{--} &      45.4&3.4 \\ 
PACS  100\,$\mu$m  &  98.3&8.5 &  82.4&4.5  \\ 
PACS  160\,$\mu$m  & 169.6&11.1 & 153.0&9.0 \\ 
SPIRE 250\,$\mu$m  & 123.3&13.4 & 110.7 &  25.2  \\ 
SPIRE 350\,$\mu$m  &  53.8&18.1 &  69.3 &  22.8 \\
SPIRE 500\,$\mu$m  &  \multicolumn{2}{c}{$<$57.3 (3$\sigma$)} & \multicolumn{2}{c}{$<$60 (3$\sigma$)} \\ \hline
APEX 350\,$\mu$m  & 58 & 13 & \multicolumn{2}{c}{}  & \citet{Lakicevic:2012wb} \\
ALMA 450\,$\mu$m &  \multicolumn{2}{c}{} &45 & 15  & \citet{Zanardo:2014}\\
ALMA 870\,$\mu$m &  \multicolumn{2}{c}{} &4.9 & 1.6  & \citet{Zanardo:2014} \\
\hline 
\end{tabular}\\
\end{center}
\end{table}
%_________________________________________________________________

%====================
%===============

\subsection{The SPIRE Fourier Transform spectrum}

A 447--1540\,GHz SPIRE Fourier Transform spectrum of SN\,1987A was 
obtained on 2012 June 12th (day 9241; OBSID 1342246989) in a {\em 
Herschel} guaranteed time program (GT2$_{-}$mbarlow$_{-}$1) with a total 
duration of 14156\,sec, and a spectral resolution of 1.2\,GHz. The 
spectrum was reduced in HIPE v11, and part of the spectrum was presented 
by \citet{Kamenetzky:2013fv}. For the $J$=6--5 and 7--6 CO lines the 
measured line fluxes were $(7\pm3)\times10^{-18}$ and 
$(8\pm2)\times10^{-18}$\,W\,m$^{-2}$. Together with ALMA data, 
\citet{Kamenetzky:2013fv} estimated a CO temperature between 13~K and 
132\,K and a minimum CO mass of 0.01\,\Msun.

We calculated the maximum possible CO contribution to the Herschel 
photometric bands, adopting an excitation temperature of 132\,K, using an 
LTE code \citep{Matsuura:2002dg}. An upper limit of $<$9\,\% (3$\sigma$) 
is estimated for the CO line contribution to the in-band SPIRE 350\,$\mu$m 
flux, with other bands having $<3\,\%$. Hence we ignore contamination by 
CO lines of the broad-band photometric points.

\subsection{PACS observations of the [O~{\sc i}] 63\,$\mu$m line region} \label{OI}

Targeting the [O~{\sc i}] 63\,$\mu$m line, a PACS 59--70\,$\mu$m spectrum 
of SN\,1987A was obtained on 2012 October 23rd (day 9374; OBSID 
1342237430), as part of the MESS guaranteed time key program 
\citep{Groenewegen:2011p29486}, using chopped-nodded PACS range 
spectroscopy modes \citep{Poglitsch:2010bm}. The spectral resolution was 
80\,km\,s$^{-1}$ (R=1500). The total duration was 3143\,s for four 
scan-repetitions.

\begin{figure}
\centering
\resizebox{\hsize}{!}{\includegraphics*{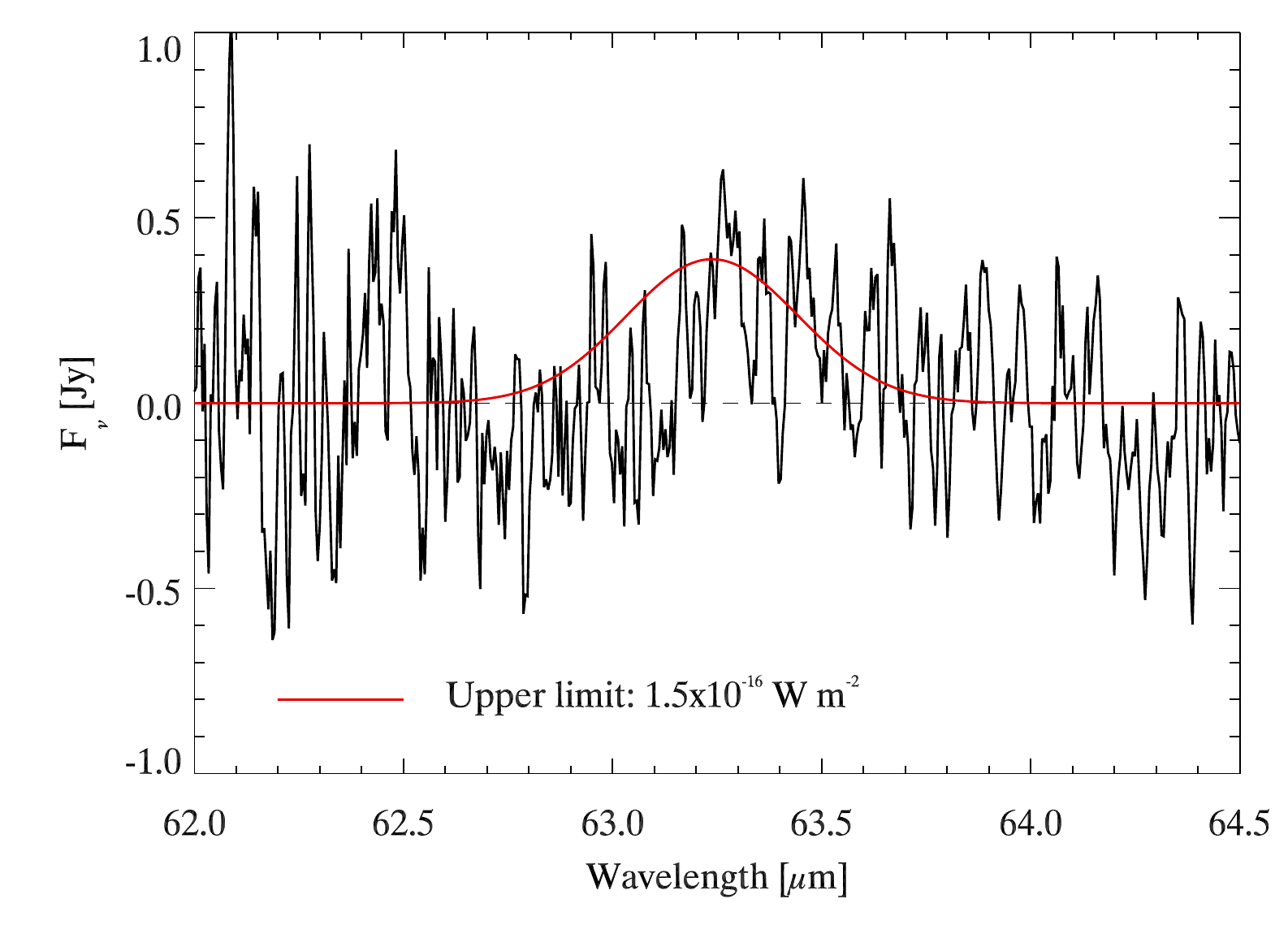}}
\caption{The 2012 PACS spectrum of SN\,1987A showing 
the region of the 63\,$\mu$m [O~{\sc i}] line, which is not detected. The 
continuous line corresponds to our 3\,$\sigma$ line flux upper limit. }
\label{fig-OI}
\end{figure}

Fig.\,\ref{fig-OI} shows the PACS spectrum and the non-detection of the 
[O~{\sc i}] 63\,$\mu$m line. An upper limit of 
$<1.5\times10^{-16}$\,W\,m$^{-2}$ is obtained for the case of a 
2300\,km\,s$^{-1}$ FWHM ejecta line width \citep{Kjr:2010p29878}. The 
upper limit for the [O~{\sc i}] line flux is consistent with model 
predictions, where the models of \citet{Kozma:1998p29912} and 
\citet{Groningsson:2008fm} predict [O~{\sc i}] line intensities of 
(0.5--1)$\times10^{-16}$\,W\,m$^{-2}$, while \citet{Jerkstrand:2011fz} 
predict $1.3 \times10^{-16}$\,W\,m$^{-2}$ from the ejecta in 2012.

For line emission originating from the ring, an upper limit of 
$<2\times10^{-17}$\,W\,m$^{-2}$ was obtained, assuming a line width of 
350\,km\,s$^{-1}$ \citep[e.g.][]{Groningsson:2008fm}. This is within the 
range expected ($<5\times10^{-18}$\,W\,m$^{-2}$) from adopting the 
predicted 63-$\mu$m/6300-\AA\ line ratio of
\citet{Groningsson:2008fm}, and the observed [O~{\sc i}] 6300-\AA\ flux 
decrease with time (Migotto et al., in preparation).

%====================
%====================

\section{Analysis}

Figure\,\ref{fig-sed} shows the SED of SN\,1987A. The newly obtained 
photometric point at 70\,$\mu$m clearly shows the dip between two discrete 
components of thermal dust emission (warm and cold). These components 
arise from two different locations within the system: the cold dust is 
located in the ejecta \citep{Indebetouw:2013vv}, while the warm dust is 
emitted in/near the equatorial ring associated with circumstellar material 
from the progenitor star \citep{Bouchet:2006p2168}.

We fitted the cold component of the thermal dust emission in order to 
estimate the required temperatures and dust masses for different grain 
compositions. The fluxes from 100--870\,$\mu$m in 2012 can be fitted 
with dust having a single temperature and a single composition 
(Fig.\,\ref{fig-sed}). The fitting was optimised using the $\chi^2$ 
minimum fitting code {\sc mpfit} in {\sc idl}.

Dust optical constants were taken from \citet{Zubko:1996p29442} for 
amorphous carbon (ACAR), from \citet{Jager:2003gl} for silicate, and from 
\citet{Begemann:1994p29483, Begemann:1997p22124} for aluminium oxide 
(Al$_2$O$_3$)  and sulphides (Mg$_{0.9}$Fe$_{0.1}$S and FeS). 
Unfortunately, the wavelength coverage of the Al$_2$O$_3$ data is too 
short ($<$200\,$\mu$m) and the Mg$_{0.9}$Fe$_{0.1}$S data are noisy beyond 
150\,$\mu$m, so that these analyses are for general guidance only. Grain 
densities ($\rho$) of 1.81, 3.3, 3.2 and 4.83\,g\,cm$^{-3}$ were adopted 
for amorphous carbon, silicates, Al$_2$O$_3$ and FeS, respectively 
\citep{Zubko:2004p13575, Draine:1984p25590, Begemann:1997p22124, 
Semenov:2003p29439}. Mg$_{0.9}$Fe$_{0.1}$S is a similar material to MgS, 
with a grain density of 2.84\,g\,cm$^{-3}$ \citep{2014pccd.book.....G}. We 
calculated the dust emissivity ($Q_\nu$) at frequency $\nu$ using Mie 
theory, and with dust mass absorption coefficients 
($\kappa_\nu=\frac{3Q_\nu}{4\rho a_d}$).  
%\,cm$^2$\,g$^{-1}$ 
A grain size, $a_d$ of 0.1\,$\mu$m is assumed. However, $\kappa_\nu$ and thus the 
inferred dust mass, does not depend on $a$ up to $a$=0.5\,$\mu$m. The only 
exception is iron, for which we have used $a$-dependent iron emissivities 
\citep{Nozawa:2006p29444}.

Our thermal dust emission calculations consider both optically 
thin and thick cases. In the optically thin case, the flux density 
$F_{\nu}$ from a dust mass ($M_d$) is given by $ F_{\nu} = M_{d} \frac{4 
\kappa_\nu \pi B_{\nu}}{4 \pi d^2}$ \citep{Hildebrand:1983tm}, where $d$ 
is the distance to the LMC, adopted to be 50\,kpc. 
In order to deal with the optically thick 
cases, we use the escape probability ($P_{\nu}$) for a photon emitted in a 
sphere \citep{2006agna.book.....O}. 
The optical depth ($\tau_\nu$) for a 
sphere was calculated using the equation $ \tau_0 (\nu) = \frac{3}{4} 
\frac{M_d}{\pi R^2} \kappa_\nu$, where $\tau_0 (\nu)$ is the radial 
optical depth along the line of sight, and a radius ($R$)  of 
$1\times10^{17}$\,cm was assumed \citep{Indebetouw:2013vv}.
The flux density is given by $ F_{\nu}' =  F_{\nu} P_{\nu}$.

Our dust model fitting results are shown in Figure~4 and summarised in 
Table\,\ref{table-dust}. Models (a)--(j) omitted the 70-$\mu$m flux point 
from the fits, while models (k)--(n) included it. For amorphous 
carbon the single temperature component fitting yielded a dust mass of 
0.5$\pm$0.1\,\Msun, with little difference in the inferred 
dust masses between the optically thick and optically thin cases 
(models a and b).
This is 
because the emission is largely optically thin, being marginally optically 
thick at 100\,$\mu$m ($\tau_{100\,\mu m}=1.2$). Fitting with silicates 
requires the use of escape probabilities, as $\tau>1$ at wavelengths
$\leqslant$160\,$\mu$m.

Overall, our fitted dust models produce dust masses and temperatures 
consistent with the previous {\em Herschel} analysis for amorphous carbon 
and iron \citep{Matsuura:2011ij}. A slight difference is found in the dust 
temperature inferred for silicates, 
partly because the higher signal-to-noise  2012 PACS photometer fluxes are slightly different 
from those measured in 2010, 
partly because different dust optical 
constants are used, and partly because previously an optically thin case 
was assumed, which is now found not to be valid at 100 and 160\,$\mu$m.

 The $\chi^2$ values are the lowest for amorphous carbon,
 suggesting that this dust species could be  the major component of the ejecta dust.
 Silicates, whose emissivities decline more steeply towards longer wavelengths
 than amorphous carbon, show larger $\chi^2$ values than amorphous carbon.

We could also fit the SED using a combination of silicates and amorphous 
carbon (Figure~4d); the combination required 0.5\,\Msun\, of silicate and 
0.3\,\Msun\, of amorphous carbon, for a combined dust mass of 
0.8~M$_\odot$ (Table~2).

The fits discussed so far did not include the 70-$\mu$m flux measured in 
2012. This flux has potential contaminants. Our {\em Herschel} PACS 
spectrum shows that the [O~{\sc i}] 63-$\mu$m line could contribute up to 
24\,\% of the 70-$\mu$m in-band flux (Sect.\ref{OI}). Further, the 
contribution from `warm' dust located in the equatorial ring was estimated 
to be about $\sim12$\,\% of the in-band flux (Sect.\,\ref{photometry}), 
corresponding to a combined contribution of up to 36\,\% of the 70-$\mu$m 
in-band flux, unrelated to the thermal emission of cold dust.
We fitted the far-infrared SED with cold dust emission, including the 
70-$\mu$m point in the fit, after subtracting off 36\,\% of the 70-$\mu$m 
in-band flux. First, we fitted with a single dust component. The fit
used amorphous carbon and the optically thick assumption was applied. The 
result is shown in Fig.~4k, with a dust temperature of 23.7$\pm$0.4\,K and 
a dust mass of 0.4$\pm$0.04\,\Msun. This dust mass and its temperature 
are nearly consistent within the 1-$\sigma$ uncertainties with the fit 
that omitted the 70-$\mu$m flux point (Table~2), but the fit is poor at 
70~$\mu$m  and noticeably worse at 250 and 350~$\mu$m, compared to the 
single component fit that omitted the 70-$\mu$m point (Fig.~4b).

 We therefore also produced fits to the entire 70--870\,$\mu$m SED using 
two dust components. The results are shown in Fig.~4k--Fig.~4m and the 
parameters are listed in Table~2. The fitting procedure found a solution 
with a warm silicate component having a temperature of 134\,K and a dust 
mass of 6$\times10^{-5}$~M$_\odot$, along with a cold amorphous carbon 
component having a temperature of 24\,K and a dust mass of 
0.5~M$_\odot$. Since the temperature of the warm component is close to 
that discussed earlier for the equatorial ring dust, this suggests that 
36\% represents an underestimate of the combined contribution to the 
70-$\mu$m in-band flux made by the 63-$\mu$m line and by the warm ring 
dust. The cold AC dust component in the 70--870-$\mu$m one and two-component 
fit has a consistent dust mass for amorphous carbon for the single-component 100--870-$\mu$m 
fit.

A possible way to reduce the required mass of silicate dust is to assume 
that the dust grains are elongated,  which can enhance their far-infrared emissivity.
For ellipsoid grains with three axes ($a$, $b$, $c$) with
$a=b\ll c \ll \lambda$, ,where $\lambda$ is the emitting wavelength \citep{Hoyle1991},
we find that for silicates the value of $\kappa$
is enhanced by a constant factor of about 2 
for wavelengths longer than $\sim$50\,$\mu$m.
The value of $\kappa$ for elongated amorphous carbon grains is enhanced by a factor of 
$\sim20$--200  from $\sim50$ to 1000\,$\mu$m. 
However, the resulting wavelength dependence of $\kappa$ is considerably flattened, 
providing a very poor fit to the Herschel and ALMA data. 
For silicates, 
the dust mass in SN1987A can therefore be 
0.8\,\Msun, if the nucleating silicates attain an elongated shape,  
but the emission is unlikely to be due to elongated amorphous carbon grains.
Grains nucleating in a radioactive environment may acquire an electric charge 
that can affect the formation and growth of these dust grains. 
This effect may preferentially affect the silicate grains: 
 because of their dielectric nature, which may cause them to attain 
an asymmetrical charge distribution; and 
 because they nucleate in an environment that is more directly exposed to hard radiation.
The presence of non-spherical grains in dense regions of the 
ISM has been inferred from observations of linearly polarised thermal dust emission 
at sub millimetre wavelengths  \citep{Hildebrand:1995kj}.
The possible presence SN-condensed  elongated silicates in the diffuse ISM 
may require modification to existing interstellar dust models 
\citep{Zubko:2004p13575, Draine:2007p17122}.

%The largest uncertainty in the derived dust masses is associated with the 
%dust properties, particularly the grain morphology.  Mie theory assumes 
%spherical grains, but astronomical grains are probably not perfect 
%spheres. An alternative choice is to use the continuous distribution of 
%ellipsoids (CDE) method, which assumes the presence of all shapes of 
%ellipsoids \citep{Bohren1983}. Elongated dust grains can increase $\kappa$ 
%at far-infrared wavelengths, decreasing the inferred dust mass of 
%SN\,1987A to 0.05\,\Msun\, for amorphous carbon, and 1.4\,\Msun\, for 
%silicate with the CDE assumption. However, the CDE assumption includes 
%extreme shapes, such as indefinitely long needles and disks, which cause 
%$\kappa$ to be enhanced but beyond the Rayleigh limit 
%($\frac{2\pi a}{\lambda} \ll 1$), and such shapes seem unlikely to be present in 
%cosmic dust \citep[see][]{Li:2008dl}. The $\chi^2$ values worsen for CDE 
%amorphous carbon fits to SN\,1987A's SED, due to their altered emissivity 
%slopes. We prefer the Mie assumption, but the inferred dust mass could be 
%reduced to some degree by mixing in some non-spherical grains.

%_________________________________________________________________
\begin{table*}
% \centering
  \caption{   Summary of the derived dust masses, and dust mass 
constraints from predicted available elemental masses \label{table-dust}}
\begin{center}
 \begin{tabular}{cllll rrrrrrl}
\hline
    & Model                  &  $P_{\nu}$ &$M_d$ (\Msun)                & $T_d$ (K)          & Reduced $\chi^2$ & $M_m$ (\Msun) \\ \hline
(a) & Amorphous carbon (AC)  & ---        & 0.5$\pm$0.1                 & 20.3$\pm$0.5       & 0.14         & 0.25\\
(b) & AC                     & $P_{\nu}$  & 0.5$\pm$0.1                 & 23.2$\pm$0.5       & 0.13         & 0.25 \\
(c) & Silicate (sil)         & $P_{\nu}$  & 2.4$\pm$0.5                 & 22.5$\pm$0.3       & 0.77         & $\sim$0.5 \\
(d) & AC + silicate          & $P_{\nu}$  & 0.5 (AC)  + 0.07 (sil)      & 23 (AC) + 22 (sil) & 0.13 \\
(e) & AC + silicate          & $P_{\nu}$  & 0.3 (AC)  + 0.5 (sil) fixed & 25 (AC) + 20 (sil) & 1.15 \\ 
(f) & Fe ($a_d=$0.5\,$\mu$m)   & $P_{\nu}$  & 0.37$\pm$0.04               & 26.9$\pm$0.7       & 1.57         & 0.24\\ 
(g) & FeS                    & ---        & 0.9$\pm$0.1                 & 31$\pm$0.2         & 0.86         & 0.24 \\ \hline
\multicolumn{5}{l}{Less important for far-infrared emission} \\ 
(h) & Fe ($a_d=$0.05\,$\mu$m)  & $P_{\nu}$  & 12.8$\pm$2.0                & 22.8$\pm$0.2       & 0.44         & 0.24\\
(i) & Mg$_{0.9}$Fe$_{0.1}$S  & ---        & 1.4                         & 19.5               & 6.33          & 0.25 \\ 
(j) & Al$_2$O$_3$            & ---        & 0.7                         & 20                 &              & 0.02\\  \hline
\multicolumn{5}{l}{Testing fits to 70\,--870$\mu$m flux, after  36\,\% of the 70\,$\mu$m flux was subtracted } \\ 
(k) & AC                     & $P_{\nu}$  & 0.4$\pm$0.04                & 23.7$\pm$0.4       & 0.96 \\
%(l) & AC + AC                & ---        & 0.7 (cold) + 0.03 (warm) & 18 + 28            & 5.89 \\
(l) & AC + AC                & $P_{\nu}$  & 0.5 (cold) + $5\times10^{-4}$ (warm) & 24 + 83  & 0.12\\
(m) & AC + Silicate          & $P_{\nu}$ (AC only)  & 0.5 (AC) + $6\times10^{-5}$ (sil)    & 23 (AC) + 134 (sil) & 0.11 \\
\hline
%\multicolumn{3}{l}{$a$-dependence} \\ 
%AC ($a$=2.0\,$\mu$m)   & thin & 0.6 & 19.7  \\
%AC ($a$=5.0\,$\mu$m)   & thin & 0.7 & 17.3 \\ \hline
%AC ($a$=10\,$\mu$m)    & thin & 0.3 & 16.6 \\
%AC ($a$=20\,$\mu$m)    & thin & 0.02 & 27.8 \\ 
\end{tabular}\\
$P_{\nu}$ shows whether escape probabilities were involved in the calculation.
$M_d$: dust mass, $T_d$: dust temperature and $a_d$: grain radius, which is 0.1\,$\mu$m, unless stated.
$M_m$: maximum dust mass allowed by elemental masses predicted by current explosive nucleosynthesis models \citep{Rauscher:2002p29812}.
Fe elemental mass includes decay from $^{56}$Ni to $^{56}$Co, then to $^{56}$Fe with half lives of 6 and 77 days.
Model (e) used fixed dust masses.
\end{center}
\vspace{0.5cm}
\end{table*}
%_________________________________________________________________

%________________________________________________________________
\begin{figure*} % Fig. 4
 \rotatebox{270}{ 
 \begin{minipage} {13cm} 
\resizebox{\hsize}{!}{\includegraphics*{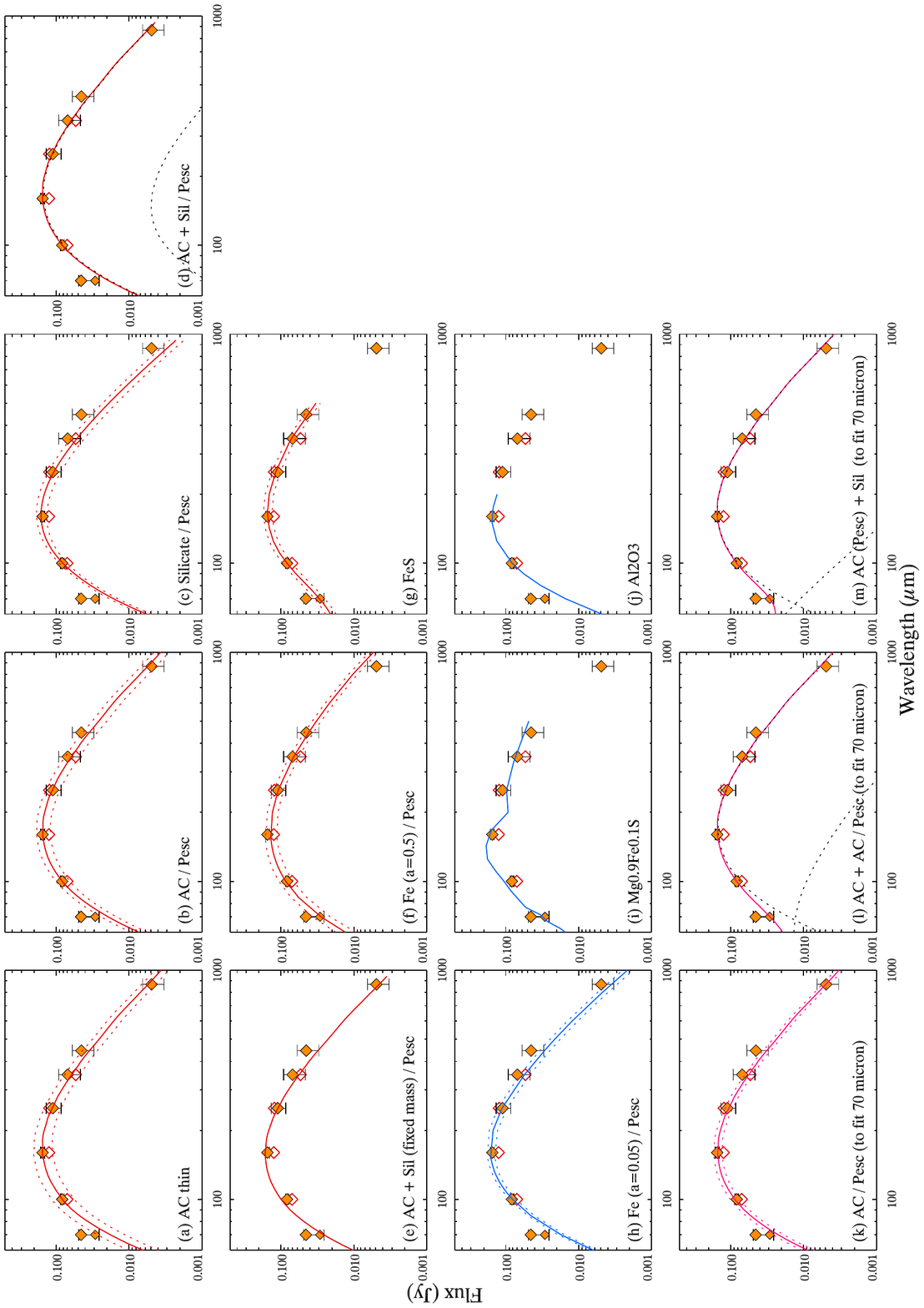}}
 \end{minipage}
 }
  \caption{The model SED fits. Diamonds show the observed fluxes in 2012, as 
in Fig.\,\ref{fig-sed}. The lower error bar limit and smaller diamonds of the 70\,$\mu$m flux includes
estimated contribution from [O~{\sc i}] and ring dust;
the 70\,$\mu$m flux with these contributions subtracted is indicated in 
filled orange diamond.
Open diamonds show the measured fluxes in 2010.
The fits are plotted as solid curves, and their uncertainty ranges 
are shown as dotted curves. The fit for Al$_2$O$_3$ in box (i) stops
at 200~$\mu$m, the longest wavelength for which optical constants are
available (see text). The
parameters for each of the fits can be found in Table\,\ref{table-dust}.
%In panel (n), silicate emission lies outside of the plotted range.
The label $P_{\rm esc}$ indicates that the fitting involve escape probabilities.
The models (h--j), plotted in blue lines, have fitted dust mass
a factor of five larger dust mass than dust mass constraints from predicted by nuclear synthesis models predicted.
      \label{plot_fit}}
\end{figure*}
%________________________________________________________________

\section{Discussion}

Our single component model fits to the far-infrared SED with amorphous 
carbon give a dust mass of 0.5$\pm$0.1\,\Msun (Table\,\ref{table-dust}), 
while the fit with a mixture of amorphous carbon and silicates requires 
0.3~M$_\odot$ of carbon. These masses of amorphous carbon are higher than 
the mass of carbon (0.25\,\Msun) currently predicted by explosive 
nucleosynthesis models for a 19\,\Msun\, star \citep{Rauscher:2002p29812}, 
apparently implying a deficit of carbon to account for the far-infrared 
emission, although the carbon deficit is small (0.05~M$_\odot$) in the 
case of the mixed AC+silicate model. We note that that the C/O ratios 
predicted by nucleosynthesis models for core-collapse supernovae (CCSNe) 
may not be accurate for all initial mass cases. For example, the Crab 
Nebula progenitor star has been estimated to have had an initial main 
sequence mass of 9--13~M$_\odot$ 
\citep{Hester:2008kx, Smith:2013fu}. CCSN yield 
predictions for 11--13~M$_\odot$ initial mass models, the lowest masses for 
which nucleosynthesis predictions are currently available, all predict C/O 
mass ratios of much less than unity 
\citep{Woosley:1995jn, Thielemann:1996dw, Nomoto:2006di},
as do higher mass models, whereas the photionization 
modelling analysis by 
\citet{MacAlpine:2008kn}
of many locations in the Crab 
Nebula found the nebula to be overwhelmingly carbon-rich (C/O$>$1, 
both by number and by mass).

Although a 10-$\mu$m or 18-$\mu$m silicate emission or absorption feature 
was not seen during the first three years of SN~1987A's evolution, when 
its dust SED peaked at mid-infrared wavelengths \citep{Wooden:1993p29432}, 
indications that silicate dust may have formed as well as carbon dust comes 
from the fact that molecular SiO vibrational emission was detected at 
early times \citep[since day 164; ][]{Roche:1991vv}, but disappeared from 
mid-infrared spectra at about the time that dust formation began 
\citep{Wooden:1993p29432}, potentially due to depletion caused by silicate 
dust formation \citep{Sarangi:2013bj}. Recent ALMA submillimeter 
observations of SN~1987A have also detected SiO, via its rotational 
emission spectrum \citep{Kamenetzky:2013fv}. The current SED can be fitted 
with 2.9$\pm$0.5\,\Msun\, of silicate grains alone (Fig.~4c), but this 
exceeds by a factor of six the silicate mass limit ($\sim$0.5\,\Msun\,) 
from the Si, Fe and Mg abundances predicted by explosive nucleosynthesis 
models \citep{Thielemann:1990p29472, Rauscher:2002p29812}. Hence, it seems 
implausible for only silicate dust to be present in the ejecta. We have 
shown however that a combination of 0.3\,\Msun\, of amorphous carbon and 
0.5\,\Msun\, of silicates, for a total dust mass of 0.8\,\Msun, can fit 
the observed SED while satisfying the currently predicted elemental 
abundance limits for silicates and is close to the currently predicted 
elemental mass limit for carbon. If one disregards the predicted elemental 
carbon mass limit of 0.25~M$_\odot$, then the lowest total dust mass that 
can fit the current SED of SN~1987A is 0.5$\pm$0.1~M$_\odot$ of amorphous 
carbon.

The value of $\kappa$ for iron grains has a steep dependence on grain 
radius $a_d$. For $a_d=$0.5\,$\mu$m, the derived dust mass is 0.3\,\Msun, but 
it increases to 14\,\Msun\, for $a_d=0.05$\,$\mu$m 
(Table\,\ref{table-dust}). As it is unlikely for all iron grains to have a 
radius of 0.5\,$\mu$m, we consider that iron grains are not the major 
source of the far-infrared emission.

Although \citet{Sarangi:2013bj} predicted Al$_2$O$_3$ to be the species 
with the largest dust mass fraction for a star of initial mass 19\,\Msun\, 
their predicted Al$_2$O$_3$ mass of 0.02\,\Msun\, is insufficient to 
explain the observed far-infrared emission. Instead, sulphides, e.g. 
FeS and Mg$_{0.9}$Fe$_{0.1}$S, in addition to amorphous carbon and 
silicates, could potentially contribute to the far-infrared emission.

Within the ejecta of a core-collapse supernova it should be feasible to 
form a mixture of different dust species. The immediate progenitor star 
will have been composed of multiple layers of different elements, 
resulting from the sequence of nuclear reactions. One layer has more 
carbon than oxygen atoms, while a silicon-rich layer also contains oxygen 
and iron \citep[e.g.][]{Rauscher:2002p29812}. It is believed that 
different layers are largely unmixed after the SN explosion. Consequently, 
amorphous carbon dust and silicates can be formed from material 
originating from the different layers. Given that multiple types of 
silicates have been inferred to be present in the Galactic supernova 
remnant Cassiopeia A \citep{Rho:2008p414, Arendt:2014ka} and that chemical models predict 
multiple dust species \citep{Sarangi:2013bj}, it seems plausible that SN 
1987A will have a mixture of different dust species, including both 
carbonaceous and silicaceous grains.

The mid-infrared observations of SN 1987A at early times ($<$ 1000 days) 
implied much smaller ejecta dust masses than derived from the {\em 
Herschel} measurements, e.g.  \citet{Wooden:1993p29432} estimated a dust 
mass of $\sim10^{-4}$\,\Msun, and later radiative transfer analyses have 
confirmed that there was $<$few$\times10^{-3}$\,\Msun\, of dust at those 
epochs \citep{Ercolano:2007p2875, wesson2014}. The large difference 
between the ejecta dust masses measured then and now implies that the dust 
mass must have increased significantly over the last 20 years.

The processes to form such a large dust mass over 20\,years may involve 
dust grain growth. For overall C/O number ratios of less than unity, the 
chemical models of \citet{Sarangi:2013bj} 
predict that carbon atoms should be primarily locked up in CO molecules at 
early times ($<$1000\,days), preventing the formation of a large mass of 
amorphous carbon (at most $5.5\times10^{-3}$\,\Msun). 
\citet{Clayton:2011fp} suggested that a possible solution to forming 
amorphous carbon is the dissociation of CO by energetic electrons created 
by Compton scattering of $\gamma$-rays from radioactive decays. As the 
ejecta expands and the gas density reduces, the shielding of electrons 
could decrease, potentially making CO dissociation more efficient. 
Additionally, a small fraction of the X-ray radiation from the ring 
\citep{Helder:2013dv} could potentially penetrate into the clumpy ejecta, 
dissociating CO \citep{Hollenbach:1997di}. The dissociation rate of CO 
depends heavily on the gas density and extinction,
increasing at lower densities. While the ejecta expands, the gas density 
decreases, thus the CO dissociation rate could have increased with time. 
If atoms or molecules accrete onto existing dust grains, the total 
dust mass should increase with time.

%It seems likely that the ejecta has remained chemically active over the 
%last 25 years.  For 1.8\,\Msun\ of oxygen, the most abundant element in 
%the ejecta \citep{Rauscher:2002p29812} and adopting a mean expansion 
%velocity of 2240~km~s$^{-1}$ \citep{{Kamenetzky:2013fv}, the mean oxygen 
%number density in the ejecta after 25 years would be 
%$6\times10^3$\,cm$^{-3}$. Accounting for other species would raise the
%density by a factor of a few. If the ejecta is filled with clumps 
%\citep{Lucy:1989p29407} rather than uniform, the gas density could be an 
%order of magnitude higher. For a temperature 25\,K the mean 
%thermal velocity of oxygen atoms would be $10^4$\,cm\,s$^{-1}$, leading
%to one particle colliding with another once per $\sim10^3$~seconds.

The supernova explosion produced radioactive isotopes, and the energy 
generated by their decay should serve as the main heating source of the 
ejecta. $^{44}$Ti is predicted to have 
become the main radioactive heating source several years after the explosion
\citep{Thielemann:1990p29472}. 
The deposited energy from $^{44}$Ti, extrapolated from 
\citet{Jerkstrand:2011fz}, is estimated to have been 
%$1.6\times10^{29}$ W\,m$^{-2}$ 
412\,\Lsun\, in 2012. This is sufficient to heat the ejecta 
dust grains, which have a total luminosity of 230\,\Lsun. In the SN 
system, X-ray radiation from the ring can provide a large luminosity of 
$\sim500$\,\Lsun\, \citep{Sturm:2010p29859}. However, since as seen from 
the equatorial ring, the ejecta core occupies 
only about 5\% of the sky, only a small fraction of this luminosity (about 
$\sim50$\,\Lsun) may reach the ejecta. So currently $^{44}$Ti should be 
the main dust 
heating source. The temperature of the dust in the ejecta of SN~1987A is 
found to be confined to a small range, and this may also support $^{44}$Ti 
as the main heating source for the dust in the ejecta. The 
expected diffuse 
distribution of $^{44}$Ti within the ejecta seems more likely to heat the 
dust grains relatively uniformly, whereas external X-ray heating should 
produce a dust grain temperature gradient within the ejecta.

That the temperature of dust emission from SN 1987A can be fitted by a 
single temperature component is a significant difference from 
the Crab Nebula, an older supernova remnant, 
which shows a broader range of dust 
temperatures \citep{Gomez:2012fm}. This is most likely caused by the 
different heating sources and distributions of dust with respect to the 
heating sources. While the main heating source in the ejecta of SN~1987A 
is $^{44}$Ti decays, the main dust heating source in the Crab Nebula is 
synchrotron radiation from its pulsar wind nebula \citep{temim:2013vi}.

\section{Conclusions}

We have presented dedicated {\em Herschel} observations of SN 1987A, taken 
in 2012. Our photometric and spectroscopic observations confirm that its 
far-infrared emission is dominated by thermal dust emission from the 
ejecta, 
with a minimal contribution from lines, confirming previous analyses 
that a large mass of dust has formed in the ejecta after the explosion. If
we consider predicted elemental mass limits, we would conclude that it 
is most likely to have formed 0.8\,\Msun\, dust in the ejecta, consisting 
of a combination of 0.5~M$_\odot$ of silicates and 0.3~M$_\odot$ of 
amorphous carbon. If the currently predicted nucleosynthetic limit of 
$\sim$0.25~M$_\odot$ of carbon is ignored, then the minimum dust 
mass that can fit the observed SED is 0.5~M$_\odot$ of amorphous carbon.

It will be interesting to monitor the ejecta dust in the future, when the 
main ejecta plunges into the circumstellar ring. As the ejecta expand, the 
ejecta dust should interact with reverse shocks, with the potential 
destruction of dust grains 
\citep[e.g.][]{Jones:1994p8385, Nozawa:2006p29444, Silvia:2012br}.
The efficiency of destruction depends on the 
position angle of the ejecta with respect to the ring. If most of the 
ejecta dust can survive future reverse shocks, and future ejecta-ambient 
ISM shocks, then CCSNe such as SN\,1987A may provide a major source of 
the dust found in the ISMs of galaxies 
\citep[e.g.][]{Matsuura:2009p29906, Dwek:2011p29471, Rowlands:2014dq}.

\acknowledgments

MM acknowledges support from the UK STFC (ST/J001511/1). RJI and LDunne acknowledge 
support from the European Research Council in the form of Advanced Grant 
COSMICISM. PvH acknowledges support from the Belgian Science Policy Office 
through the ESA PRODEX program.

%Facilities: \facility{Herschel Space Observatory}.

%\bibliography{sn1987a}

\end{document}